# Microstructure and phase transformation of nickel-titanium shape memory alloy fabricated by directed energy deposition with in-situ heat treatment

Shiming Gao[1], Ojo Philip Bodunde[1], Mian Qin[1], Wei-Hsin Liao[1,3,*], Ping Guo[2,*]

[1]Department of Mechanical and Automation Engineering, The Chinese University of Hong Kong, Shatin, Hong Kong, China

[2]Department of Mechanical Engineering, Northwestern University, Evanston, IL, USA

[3]Institute of Intelligent Design and Manufacturing, The Chinese University of Hong Kong, Shatin, Hong Kong, China

*Corresponding authors: Wei-Hsin Liao: whliao@cuhk.edu.hk

Ping Guo: ping.guo@northwestern.edu

**Abstract**

Additive manufacturing has been vastly applied to fabricate various structures of nickel-titanium (NiTi) shape memory alloys due to its flexibility to create complex structures with minimal defects. However, the microstructure heterogeneity and secondary phase formation are two main problems that impede the further application of NiTi alloys. Although post-heat treatment is usually adopted to improve or manipulate NiTi alloy properties, it cannot realize the spatial control of thermal and/or mechanical properties of NiTi alloys. To overcome the limitations of uniform post-heat treatment, this study proposes an in-situ heat treatment strategy that is integrated into the directed energy deposition of NiTi alloys. The proposed method will potentially lead to new manufacturing capabilities to achieve location-dependent performance or property manipulation. The influences of in-situ heat treatment on the thermal and mechanical properties of printed NiTi structures were investigated. The investigations were carried out in terms of thermal cycling, microstructure evolution, and mechanical properties by 3D finite element simulations and experimental characterizations. A low-power laser beam was adopted to localize the in-situ heat treatment only to the current printed layer, facilitating a reverse peritectic reaction and a transient high solution treatment successively. The proposed in-situ heat treatment on the specimen results in a more obvious phase transformation peak in the differential scanning calorimetry curves, about 50%~70% volume reduction for the $Ti_2Ni$ phase, and approximately 35 HV reduction on microhardness.

**Keywords:** Directed energy deposition; near-equiatomic NiTi; heat treatment; shape memory effect

## 1. Introduction

Near-equiatomic nickel-titanium (NiTi), also known as Nitinol, is one of the most widely used shape memory alloys (SMA) due to its unique functional characteristics, such as shape memory effect, super-elasticity, corrosion resistance, etc. Due to these attractive properties, NiTi alloys are used for many applications in the automotive, biomedical, and aerospace industries [1-4]. However, the manufacturing and processing of NiTi alloys are difficult since the alloy component is highly sensitive to the production environment, especially the active titanium element [5,6]. A slight variation in the composition of NiTi alloys caused by impurity inclusions and element evaporation will lead to a significant change in their final properties. The conventional machining of NiTi alloy is limited by its poor workability due to high work hardening, high strength, and susceptibility to burrs and adhesion [7,8]. Powder metallurgy and casting usually resulted in contamination, chemical segregation, porosity, and unsatisfactory functional properties [9-11]. The current manufacturing limitations in achieving part complexity and controlling microstructure distributions have impeded the further application of NiTi alloys.

The latest developments in NiTi alloy manufacturing introduce the application of additive manufacturing (AM), including selective laser melting (SLM), selective electron beam melting (SEBM), and directed energy deposition (DED), etc. These developments promise greater flexibility for the fabrication of complex structures with minimal defects, allowing tunable shape memory effect and super-elasticity. The relationship between the AM process parameters and the thermal and mechanical properties of fabricated NiTi alloys has been under systematic investigations in the past, including laser power [12-14], scanning speed [15-18], scanning strategy [19-23], hatching space [24-26], design strategy [27], and chamber oxygen level [28]. These process parameters will ultimately alter the width and position of the phase transformation peak and influence strain recoverability [29,30]. Improper processing parameters would result in undesirable properties, fabrication imperfections, or even process failure. To improve and tailor the as-deposited NiTi alloy properties, heat treatment as an effective post-process method has been adopted. Marattukalam et al. [31] investigated the effects of heat treatment at 500°C and 1000°C, respectively, on the phase transformation and corrosion behavior of laser deposited NiTi alloys. It was found that the heat treatment annihilated internal defects and decreased the forward and reverse transformation temperatures from 80–90°C to 20–40°C. A decline in the corrosion resistance after annealing was also observed. Pu et al. [32] utilized solution treatment at approximately 700°C and subsequent aging treatment to improve the uniformity of fabricated samples. The results indicated that the variances of super-elasticity and phase transformation behavior in the building direction could be eliminated. Li et al. [33] assessed the impact of homogenized heat treatment on the microstructure and phase transformation at certain temperatures between 950°C and 1000°C. In their result, the reduction of $Ti_2Ni$ through heat treatment effectively improved the microstructure uniformity, shape memory effect, and ductility of fabricated NiTi structures. Lu et al. [34] adopted a 1000°C high solution treatment method to successfully disperse nanoscale $Ti_2Ni$ precipitates throughout the grain interior. An ultra-high tensile strength of 880 ±13 MPa and an excellent shape memory effect with a recovery rate larger than 90% were demonstrated in their results. Sun et al. [35] studied the effect of annealing treatment on the transformation behaviors of samples undergoing compression. It was found that samples annealed at 300°C and 450°C exhibited one-stage phase transformation during heating and cooling, while specimens annealed at 600°C exhibited two-stage transformation on cooling and one-stage transformation on heating. However, the above-mentioned post-heat treatment methods cannot realize the spatial control of thermal and mechanical properties of as-deposited structures due to the lack of control of thermal distribution in uniform post-heat treatment.

The remelting method is usually adopted in AM processes to improve deposition qualities, such as surface roughness, relative density, residual stress, etc. Brodie et al. [36] integrated a 'remelting' scan strategy into the SLM process to improve the melting of Ta powders. Keyhole formation was effectively avoided, while the yield strength was increased from 426 ±15 MPa to 545±9 MPa. Feng et al. [37] employed laser remelting to relieve thermal stress formed in the yttria-stabilized zirconia (YSZ) thermal barrier coatings (TBCs). In their experiments, the residual compressive stress was reduced by nearly 75% after laser remelting. Han et al. [38] utilized laser surface remelting to process the fabricated AlSi10Mg specimens to achieve a significant reduction in surface roughness. The corresponding microhardness exhibited an approximately 19.5% increase after laser surface remelting due to the formed finer microstructures. Gustmann et al. [39] demonstrated that an additional remelting step in the fabrication of 81.95Cu-11.85Al-3.2Ni-3Mn shape memory alloy could improve the part's relative density from 98.9% to 99.5%. The microstructures and transformation temperatures were tuned by the remelting parameters, so that transformation temperature was successfully adjusted in the range of 90°C–120°C. Bayati et al.

[40] evaluated the effects of remelting on the thermal and mechanical properties of as-fabricated NiTi parts through SLM. It was found that a proper combination of remelting parameters helped to improve density, reduce defects, and improve surface roughness, while the transformation temperatures and compression behaviors were insensitive to the remelting procedure. Previous results indicate that the remelting method is an effective way to eliminate fabrication defects as well as to tailor thermal and mechanical properties in the deposition process. On the other hand, post-process remelting is prone to element loss and has limited effects on eliminating precipitations such as $Ti_2Ni$, $Ni_3Ti_4$, etc. Besides, this remelting method is not suitable for powder flow-based methods such as DED because the remelting process will inevitably lead to the melting of extra powders and part quality variations.

The properties of as-deposited NiTi shape memory alloys are highly related to the process parameters due to the involved complex physics such as melting, solidification, evaporation, etc. Such complex physics could affect the phase precipitation and microstructure formation. The lattice dislocation and secondary phases such as $Ti_2Ni$ are usually found in the fabricated structure, which is detrimental to the product performance. On the other hand, the nature of DED is similar to heat treatment due to the rapid melting and solidification of the material. The input laser power as an energy source achieves localized heat treatment. The distribution of the temperature field in the fabricated structure can be manipulated through path planning of laser input, resulting in location-dependent temperature history.

Inspired by the idea of using the laser beam in DED for localized heat treatment, we propose an in-situ heat treatment strategy to fabricate NiTi shape memory alloys in this study. Based on the binary equilibrium of the Ti-Ni phase diagram [41], the intermetallic $Ti_2Ni$ could dissolve back into the NiTi matrix at around 1000°C with a reverse peritectic reaction: $Ti_2Ni + NiTi \rightarrow L(Ti) + NiTi$. Moreover, the high solution treatment in the temperature range from 800°C to 1050°C has been proven to reduce the $Ti_2Ni$ phase and homogenize microstructures. Therefore, we propose to adjust the laser source in DED to a low-power beam to introduce an additional reheat treatment to the latest deposited layer up to the peritectic reaction temperature as a high temperature treatment. Within this temperature range, the reverse peritectic reaction and the instant high solution treatment are expected to occur successively to reduce the $Ti_2Ni$ fraction and homogenize the microstructure. The proposed localized heat treatment provides the potential to achieve spatial control of the thermal and mechanical properties of fabricated NiTi parts. In this study, the temperature history with and without heat treatment (HT) was firstly simulated and compared to reveal the possibility of reverse peritectic reaction and instant high solution treatment. The corresponding phase constituent and microstructure were then investigated to evaluate the amount of $Ti_2Ni$ reduction and microstructure variation. Finally, the forward and reversed transformation temperatures and the microhardness were measured to investigate the changes in the thermal and mechanical properties, which corroborate the variations of $Ti_2Ni$. This research provides the initial study for the in-situ heat treatment strategy of SMA in DED to manipulate the local transformation temperature and to tailor for the microstructure composition.

## 2. Strategy and experimental details
### 2.1 In-situ heat treatment strategy

The deposition process for DED without in-situ heat treatment is schematically shown in Fig. 1(a). For every single track/layer deposition, the high-power laser beam scans across the substrate and creates a molten pool on the surface. Meanwhile, the powder stock is carried by inert argon gas through a coaxial nozzle and fed into the molten pool. The molten material solidifies and forms the track after the energy deposition. The structure then is fabricated through this repeated layer-by-layer deposition. For DED with in-situ heat treatment, after a single track/layer is deposited similar to a conventional DED method, an additional low-power laser beam will repeat the trajectory to reheat the deposited layer as shown in Fig. 1(b). The reason to adopt an additional low-power laser beam is that the naturally induced heat treatment caused by subsequent printing is uncontrollable and unstable. During the deposition of the *N+1* layer, the laser power reaching the preceding *N* layer is either too strong, resulting in remelting, or too weak to reach the high solution temperature. Hence, the requirements for a reverse peritectic reaction and a high solution treatment cannot be met. The choice of the low-power laser beam should ensure that no additional powder is melted and the laser can heat the latest deposited track/layer up to 1000°C as auxiliary heat treatment. At this temperature, the current printed layer will first undergo a reverse peritectic reaction: $Ti_2Ni + NiTi \rightarrow L(Ti) + NiTi$, because the melting temperature of $Ti_2Ni$ is around 984°C. As the temperature drops below 984°C and above 800°C, the layer then experiences an instant high solution treatment, which absorbs the remaining $Ti_2Ni$ phase into NiTi solid solution. Such two phenomena reduce the $Ti_2Ni$ phase in the structure. The whole structure is then fabricated by repeating these procedures by alternating deposition and heat treatment.

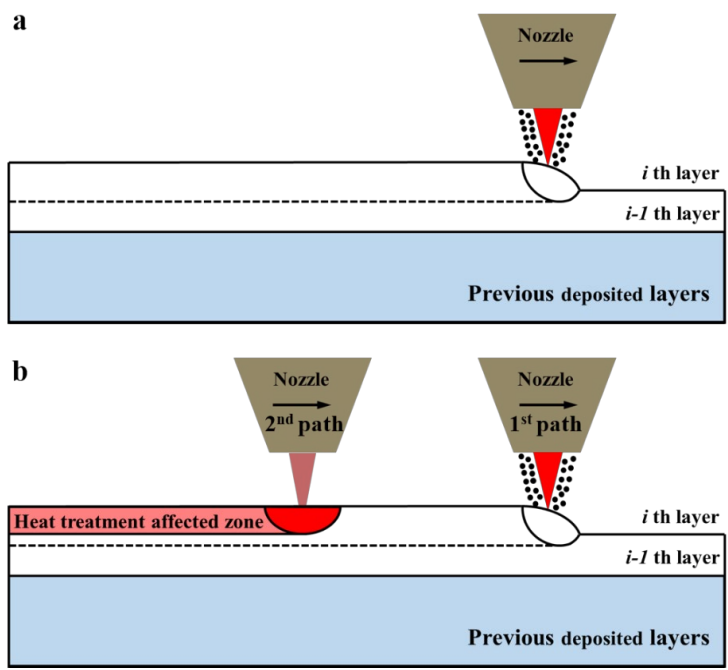

Fig. 1. Schematic of directed energy deposition: (a) without in-situ heat treatment (conventional method) and (b) with in-situ heat treatment.

**2.2 Validation method and parameter selection**

The goal of the in-situ heat treatment strategy is to achieve the spatial control of thermal and mechanical properties of fabricated NiTi specimens by manipulating the $Ti_2Ni$ dissolution and microstructure homogenization. We designed experiments to print single-wall structures with and without in-situ heat treatment, as shown in Fig. 2. The corresponding phase constituent, microstructure, and thermal and mechanical properties were correspondingly compared to verify the feasibility of in-situ heat treatment.

The fabrication process was conducted with a customized setup (see Appendix). The designed dimensions of the single-wall sample were 40 mm in length and 8 mm in height. Due to the single path printing, the width was determined by the molten pool diameter, which was affected by the laser power, scanning speed, powder feed rate, etc. The measured width of printed single-wall structures was in the range of 0.58 mm to 0.72 mm. The process schematics are demonstrated in Figs. 2(a) and 2(b). When the in-situ heat treatment was adopted, each layer was processed by two laser scans. The first scan (blue line) adopted a high laser power to achieve energy deposition (cladding). The second scan (red line) repeated the path with a low-power beam to reheat the newly formed layer as in-situ heat treatment. The detailed deposition parameters were chosen with a powder feed rate of 0.96 g/min, carriage gas flow rate of 3 L/min, protect gas flow rate of 10 L/min, laser power from 45.2 W–56.1 W, a scan speed of 2–4 mm/s, layer height of 0.03–0.05 mm. The powder flow rate remained constant during the reheating process to avoid the fluctuation in powder flow. The full experimental design and parameter combinations are listed in Table 1.

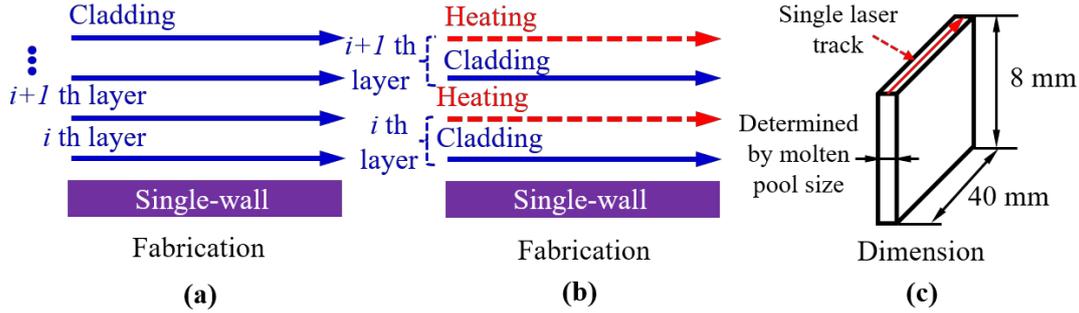

Fig. 2. Schematic of fabricated single-wall structures: (a) without in-situ heat treatment strategy, (b) with in-situ heat treatment strategy, and (c) detailed dimensions.

For the adoption of in-situ heat treatment, the low-power laser beam was generated using the same laser source for deposition. The laser power for reheating was determined based on an experimental calibration. To reach thermal equilibrium, we first printed a single-wall structure of 50 layers with each layer height of 0.05 mm (or 83 layers with each layer height of 0.03 mm). Then we adjusted the laser power to reheat the last layer so that it neither increased the layer height nor melted additional powders. The critical laser power was recorded for each parameter set listed in Table 1. The minimal value among all the recorded critical laser power at 30.5 W was adopted for the reheating treatment in the subsequent experiments. To be noted, this constant reheating laser power may only work for very simple geometry such as a single-wall structure in our case. For the in-situ heat treatment of complex geometry, the reheating laser power should be dynamically adjusted to maintain a suitable temperature range for both the reverse peritectic reaction and high solution treatment.

**Table 1**
Process parameters of DED with and without in-situ heat treatment.

| No. | Laser power, W | Scan speed, mm/s | Layer height, mm | Heat treatment power, W |
|-----|----------------|------------------|------------------|--------------------------|
| #1  | 45.2           | 2                | 0.05             | 0                        |
| #2  | 56.1           | 2                | 0.05             | 0                        |
| #3  | 45.2           | 4                | 0.03             | 0                        |
| #4  | 56.1           | 4                | 0.03             | 0                        |
| #5  | 45.2           | 2                | 0.05             | 30.5                     |
| #6  | 56.1           | 2                | 0.05             | 30.5                     |
| #7  | 45.2           | 4                | 0.03             | 30.5                     |
| #8  | 56.1           | 4                | 0.03             | 30.5                     |

## 2.3 Finite element simulation and model description

To further verify the feasibility that the in-situ heat treatment will dissolve the $Ti_2Ni$ phase and help to homogenize microstructures, a 3D finite element (FE) model was established in COMOSL 5.4 to simulate the deposition process with and without in-situ heat treatment. The corresponding FE model is shown in Fig. 3. The solid heat transfer module was adopted to deal with the temperature evolution during the deposition, involving multi-domain physical fields such as convection, conduction, radiation, and phase changes. The solid mechanics module with the "active and inactive" element activation method was utilized to handle the material addition. The whole deposition process was governed by energy conservation, as shown in Eq. (1):

$$\rho c_p^* \frac{\partial(T)}{\partial t} - \nabla \cdot (k\nabla T) = \dot{Q}(x, y, z) \tag{1}$$

where $\rho$ is the material density. $k$ is the thermal conductivity. $\dot{Q}(x, y, z)$ is the input heat source. $c_p^*$ is the modified specific heat capacity, considering the latent heat of fusion, expressed as Eq. (2):

$$c_p^* = c_p + \frac{L_m}{T_l - T_s} \tag{2}$$

where $c_p$ is the specific heat capacity. $L_m$ is the latent heat of fusion. $T_l$ and $T_s$ are the liquidus and solidus temperatures, respectively.

The boundary condition for the deposited surface and other surfaces were set to have convection and radiation losses as described in Eq. (3) [42]:

$$k(\nabla T \cdot \boldsymbol{n}) = h(T - T_0) + \varepsilon \sigma_b (T^4 - T_0^4) \tag{3}$$

where $\boldsymbol{n}$ is the normal vector. $h$ is the convection factor with 25 W/(m² *K). $\varepsilon$ is the surface radiation emissivity with a value of 0.3. $\sigma_b$ is the Stefan-Boltzmann constant defined as 5.67 x10⁻⁸ W/(m²*K⁴). $T_0$ is the room temperature.

The element activation condition was governed by Eq. (4) [43], which is highly related to the current laser position:

$$solid.node(x, y, z) = \begin{cases} active & if \ z \leq z_0 - \Delta h \ or \ z = z_0 \ and \ x \leq x_0 + L_{pool} \\ inactive & if \ z > z_0 \ or \ z = z_0 \ and \ x > x_0 + L_{pool} \end{cases} \tag{4}$$

where $solid.node(x, y, z)$ represents the status of the spatial element in the structure. $(x_0, z_0)$ are the coordinates of the current laser focus. $\Delta h$ is the single layer height. $L_{pool}$ is the molten pool length.

The "active and inactive" element activation method would change the laser action surface along with the deposition process. The usually adopted surface heat source is no longer suitable for this application. Therefore, as described in [44], the Goldak body heat source was used to represent the laser power input. The heat source model was described by Eqs. (5) and (6) due to the double-ellipsoid shapes.

$$Q_f(x, y, z) = \frac{6\sqrt{3}\eta P f_f}{abc_f \pi \sqrt{\pi}} \exp\left(-3\left(\frac{(x - x_0)^2}{c_f^2} + \frac{(y - y_0)^2}{a^2} + \frac{(z - z_0)^2}{b^2}\right)\right) \tag{5}$$

$$Q_r(x, y, z) = \frac{6\sqrt{3}\eta P f_r}{abc_r \pi \sqrt{\pi}} \exp\left(-3\left(\frac{(x - x_0)^2}{c_r^2} + \frac{(y - y_0)^2}{a^2} + \frac{(z - z_0)^2}{b^2}\right)\right) \tag{6}$$

where $P$ is the total input laser power. $\eta$ is the effective laser absorption efficiency, which considers the energy loss caused by the reflection or scattering. $Q_f(x, y, z)$ and $Q_r(x, y, z)$ are the heat flux density in the front and rear part of the source model, respectively. $f_f$ and $f_r$ are the heat fraction factors in the front and rear quadrants, where $f_f+f_r$ is equal to 2. $a$, $b$, $c_f$, $c_r$ are the width, depth, front length, and rear length of the molten pool, respectively.

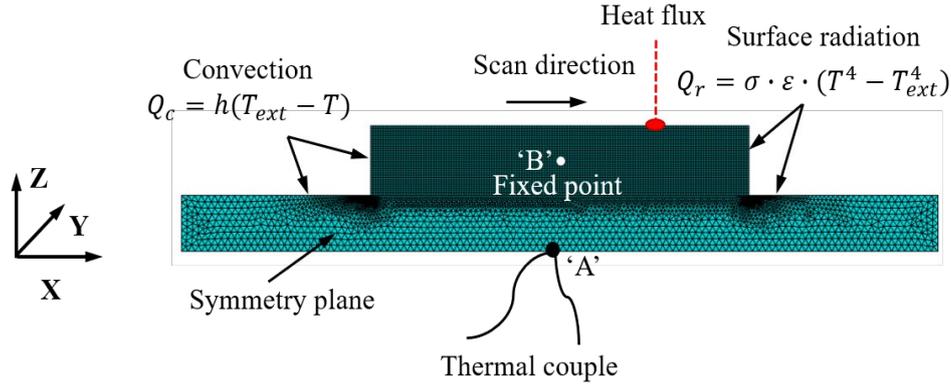

Fig. 3. Schematic of the finite element simulation model.

For high calculation efficiency, only half of the model was considered in the simulation. A symmetry boundary condition was given as follows:

$$\boldsymbol{n} \cdot (k\nabla T) = 0 \qquad (7)$$

A more detailed configuration of the FE model and simulation parameters are described in Appendix. The simulation parameters were calibrated by the temperature measurement using a thermocouple mounted on the bottom surface of the substrate. The single-wall structure was arranged to be directly deposited above the location of the thermocouple. The corresponding simulated and measured temperature histories at point 'A' on the bottom surface of the substrate, as shown in Fig. 3, were recorded and compared to calibrate the simulation parameters. To investigate whether the latest deposited layer undergoes a reverse peritectic reaction and high solution treatment under the reheating process, the temperature history of a fixed point 'B' was studied. To avoid the temperature fluctuation influenced by the cold substrate and heat accumulation, the fixed point 'B' was set at layer $N$ located in the middle of the whole FE model, as shown in Fig. 3. A state of thermal equilibrium was considered to be achieved at this position when the molten pool morphology became stable. The subsequent temperature history of this fixed point in the deposition process from layer $N$ to layer $N+5$ was then recorded and investigated to analyze the potential mechanisms of reverse peritectic reaction and microstructure homogenization.

## 2.4 Characterization

After the sample was fabricated, its middle section was cut by electrical discharge machining (EDM) along the scan direction. The composition analysis, microstructure observation, as well as thermal and mechanical tests were then performed on the cut sample. A JEOL JSM-6400 scanning electron microscope (SEM) equipped with an Oxford Instruments Energy Dispersive Spectroscopy (EDS) was adopted to investigate the microstructure morphology and elemental composition of the specimen. A Smartlab X-ray diffractometer (XRD) was utilized to study the phase constituents at room temperature. The volume fractions of the $Ti_2Ni$ phase were calculated using *ImageJ* software. The samples for microstructure observation were first polished by rough SiC sandpapers, then precisely polished using 0.5 μm diamond suspensions on synthetic cloths, and finally etched for 30 s with a mixed solution composed of 2% HF, 10% $HNO_3$, and 88% $H_2O$ in volume fraction, respectively. The martensite forward and reverse transformation temperatures were measured by Mettler Toledo differential scanning calorimetry (DSC) with a heating and cooling rate of 10°C/min from -50°C to 70°C in an argon atmosphere. The IndentaMet 1114 Vickers indenter (Vickers) was adopted to measure the microhardness with 100 g loading force and 15 s holding time. For each point, three indents were taken for repeatability.

## 3. Results and discussion
### 3.1 Temperature history

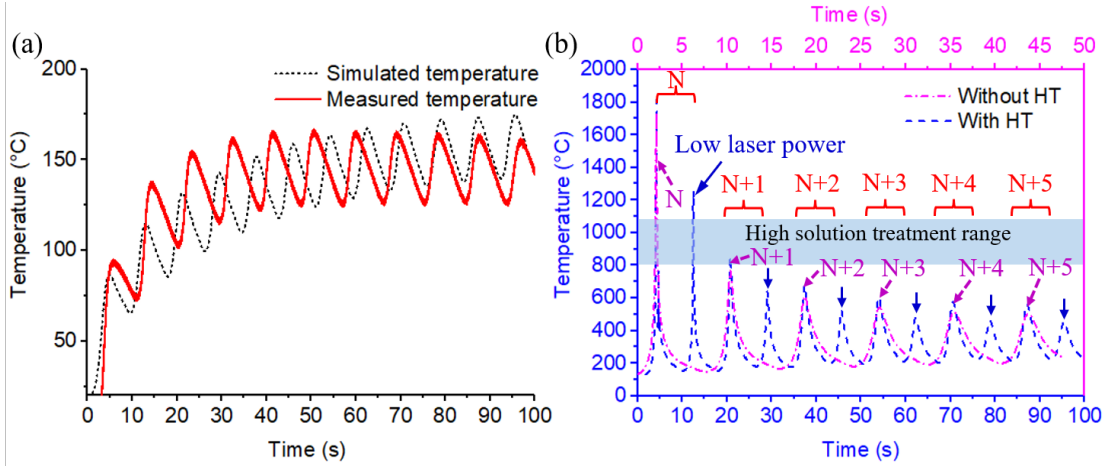

Fig. 4. (a) Simulated and measured temperature history of point 'A'; and (b) simulated thermal cycling history of point 'B' in layer $N$ during the deposition process from layer $N$ to $N$+5 layers.

Fig. 4(a) shows the simulated and measured temperature history of point 'A' to validate the accuracy of the simulation model. The temperatures were recorded in the simulation and experiments with a laser power of 45.2 W and a scan speed of 2 mm/s. A maximum error of 5.4% is observed in the simulated and experimental results. It indicates that the established model has acceptable reliability to predict the temperature history of an actual deposition process. To test whether the $Ti_2Ni$ phase will dissolve back into the NiTi matrix and whether the microstructure will be homogenized, the wall structures fabricated with and without in-situ heat treatment (HT) were then simulated according to parameters #1 and #5 listed in Table 1. Fig. 4(b) presents the simulated thermal cycling history of the fixed point 'B' when the structure was deposited from layer $N$ to layer $N$+5. Compared to the sample without HT strategy, the latest layer in the specimen fabricated with in-situ HT undergoes an additional high solution treatment in the reheating process. Meanwhile, the maximum temperature during reheating process reaches about ~1207°C, above the peritectic reaction temperature of $Ti_2Ni$ precipitation. Based on previous research [45,46], the high temperature treatment of NiTi alloy up to 1100°C can effectively decrease the grain size and volume fraction of the $Ti_2Ni$ phase. Moreover, high solution treatment has been proven to improve microstructure homogenization [47] and eliminate dislocations [48]. Therefore, it is reasonable to believe that the in-situ HT strategy may have a positive effect on the dissolution of the $Ti_2Ni$ phase back into the NiTi matrix and the homogenization of microstructures.

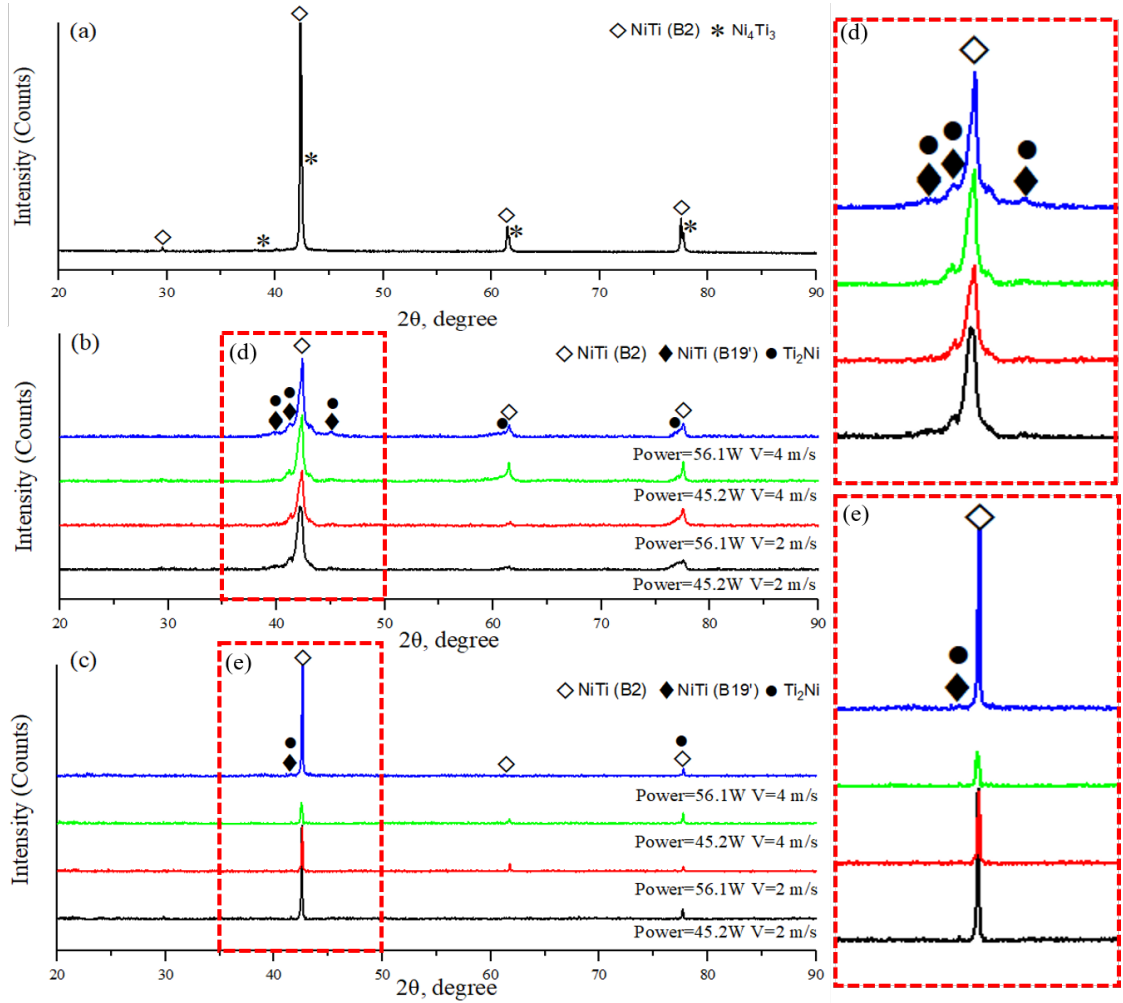

Fig. 5. XRD patterns of: (a) initial NiTi pre-alloyed powder, (b) samples fabricated without in-situ heat treatment, (c) samples fabricated with in-situ heat treatment, (d) local magnified area in (b), and (e) local magnified area in (c).

### 3.2 Phase constituent and microstructure

Fig. 5 shows the XRD analysis results of the initial powder and fabricated sample with and without in-situ heat treatment under different process parameters. As shown in Fig. 5(a), the primary peaks at 42.3°, 61.5°, and 77.4° indicate that the raw powder is mainly comprised of austenite phase (B2) at room temperature, while the minor peaks at 39.0°, 41.5°, 61.5°, and 77.4° indicate the existence of a small amount of $Ni_4Ti_3$ precipitation in the raw powders. The occurrence of $Ni_4Ti_3$ is related to the metastable phase formation during the powder preparation process, which is prone to happen with the Ni content above 50.6 at% [49]. The XRD patterns of the as-deposited sample with and without in-situ heat treatment are shown in Fig. 5(b)-(c). The austenite phase (B2), martensite phase (B19'), and $Ti_2Ni$ phase are together presented in the final structure. The occurrence of the accompanied $Ti_2Ni$ phase is related to the composition inhomogeneity during the solidification process. It has been reported that the laser burning loss [50], as well as segregation of chemical composition, will result in the local Ti-rich liquid phase in the molten pool. Therefore, a primary B2 austenite matrix with the minor irregular $Ti_2Ni$ phase is likely to be presented in the final microstructure [16, 51]. Due to the precipitation of this $Ti_2Ni$ phase, microstructure inhomogeneity and localized stress are formed. Such two phenomena impede the reversible martensite transformation propagation, resulting in the existence of the martensite phase (B19') in the XRD patterns.

Compared with the specimens fabricated without in-situ heat treatment, the samples deposited with in-situ heat treatment were observed to present weaker $Ti_2Ni$ peaks and a narrower diffraction peak width. These observations indicate the following two points. First, the fraction of the $Ti_2Ni$ phase is reduced in the measured samples, which indicates that $Ti_2Ni$ dissolved back into the NiTi matrix much faster than precipitation with in-situ HT. Similar results have also been reported when the samples were processed by high solution treatment [33]. Secondly, the narrower diffraction peak width indicates that the lattice distortion is effectively alleviated since the excessive lattice distortion will broaden the diffraction peak width due to the shift in the diffraction peaks [52].

To further verify the results of XRD observation, the corresponding SEM images were taken. The results are presented in Fig. 6 and Fig. 7. It is observed that the specimens fabricated with and without in-situ heat treatment all show two phases: the light gray (NiTi) and the dark gray ($Ti_2Ni$) phases. The detailed phase composition analysis is performed using EDS and plotted in Fig. 8(a). The atomic ratio of Ni to Ti is close to 1:2 in dark gray phases and close to 1:1 in light gray phases, representing the $Ti_2Ni$ and NiTi phases, respectively. Similar results were reported in our previous research [53] and by other researchers [54, 55]. The corresponding volume fractions of the $Ti_2Ni$ phase in the cross-section of as-deposited specimens are counted using *ImageJ* and presented in Fig. 8(b). For samples fabricated without in-situ heat treatment, when the scan speed is constant, the volume fraction of the $Ti_2Ni$ phase decreases with the increase in laser power. The higher power energy is beneficial to improve the uniformity of chemical composition in the molten pool, suppressing $Ti_2Ni$ phase precipitation. When the laser power is constant, the volume fraction of the $Ti_2Ni$ phase increases along with the increase in scan speed. This phenomenon is ascribed to the suppressed fluid convection in the molten pool. The higher scan speed accelerates the molten pool solidification. The local composition unevenness then cannot be effectively eliminated through mixing, resulting in more $Ti_2Ni$ precipitation in the structure. For the samples fabricated with in-situ heat treatment, a similar trend is also observed when the laser power is constant. However, the volume fraction of the $Ti_2Ni$ phase decreases with the laser power when the scan speed is constant at 2 mm/s but increases with the laser power when the scan speed is constant at 4 mm/s. This observation may be related to the total energy input in the deposition process. Due to the largest laser power input and the largest number of deposition layers, sample #8 has the maximum total energy input in the whole fabrication process. Its equilibrium temperature is closest to the precipitation temperature range (300°C~700°C) and the holding time is the longest, which provides more time for the precipitation of the $Ti_2Ni$ phase. Therefore, the volume fraction in sample #8 has a certain degree of increase relative to sample #7.

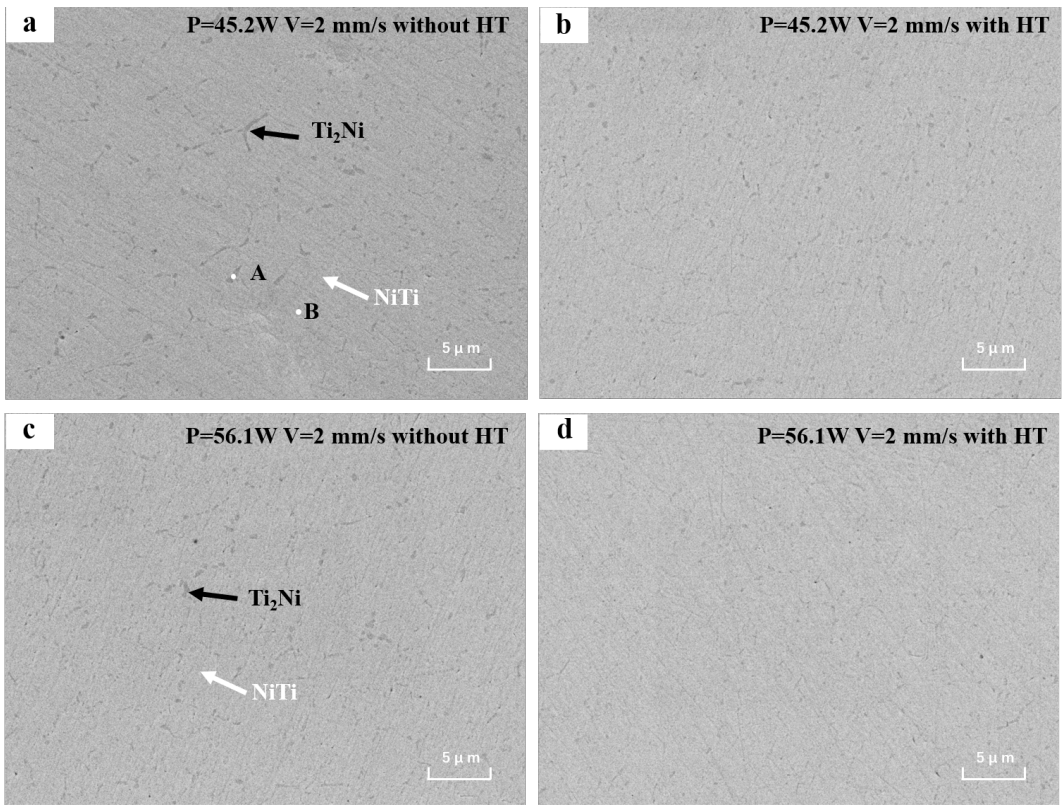

Fig. 6. Phase composition of as-deposited sample: a (#1), b (#5), c (#2), d (#6); (a, c) without in-situ heat treatment, (b, d) with in-situ heat treatment.

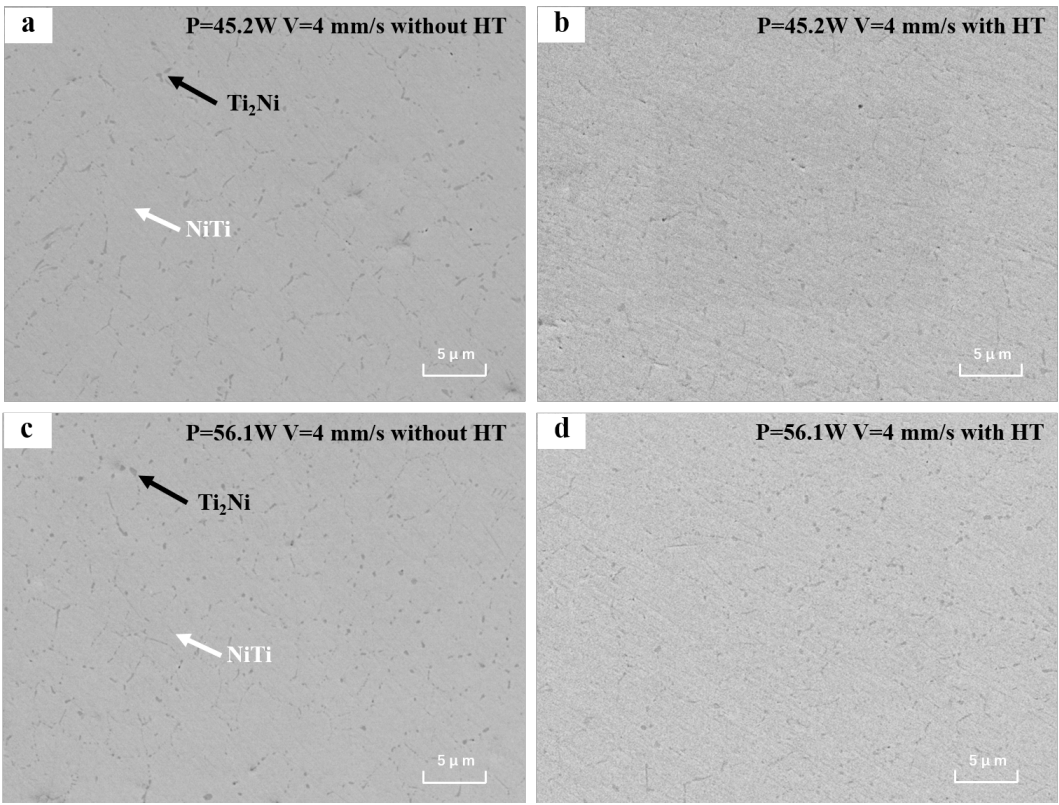

Fig. 7. Phase composition of as-deposited sample: a (#3), b (#7), c (#4), d (#8); (a, c) without in-situ heat treatment, (b, d) with in-situ heat treatment.

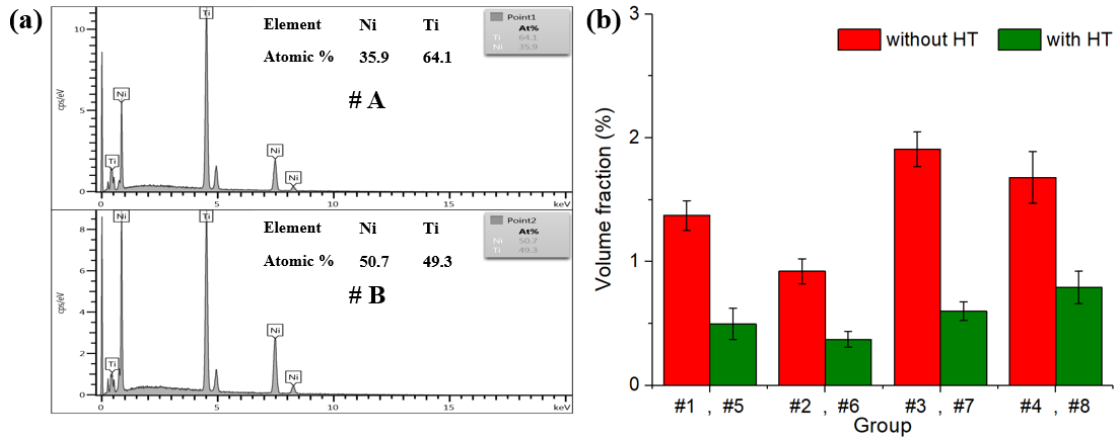

Fig. 8. (a) Element composition of point A (dark gray phase) and B (light gray phase); (b) volume fractions of $Ti_2Ni$ phase in Fig. 6 and Fig. 7.

In all cases, regardless of process parameters, the amount of $Ti_2Ni$ phase is always reduced, which indicates the effectiveness of in-situ heat treatment. The reduction of the $Ti_2Ni$ phase is about 50%~70%. This reduction of $Ti_2Ni$ precipitation is caused by the reverse peritectic reaction and/or the instant high solution treatment. A similar result was also reported by Gusev et al. [45], where the volume fraction of $Ti_2Ni$ precipitation presented a reduction trend when the NiTi alloys underwent a high temperature treatment in the range of 700°C–1100°C. Due to the reduction of the $Ti_2Ni$ fraction in the structure, the phase transformation resistance will be decreased. The smaller resistance is beneficial for improving the shape memory effect of as-deposited NiTi samples.

### 3.3 Phase transformation behavior

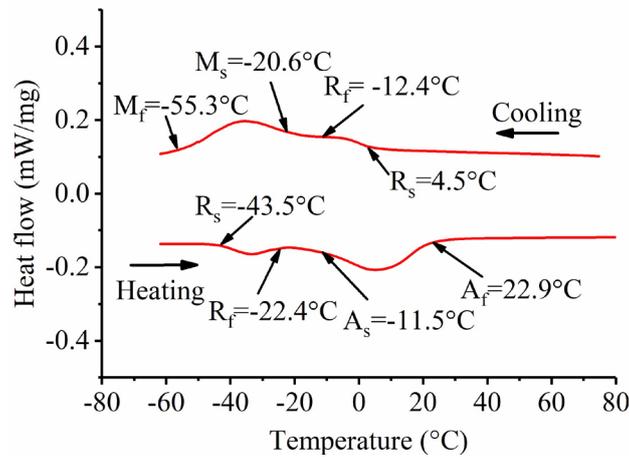

Fig. 9. DSC curves of initial powders.

The DSC curve of the raw powder is shown in Fig. 9. Six important transformation temperature points ($R_s$, $R_f$, $M_s$, $M_f$, $A_s$, $A_f$) are marked to indicate the transformation properties. The tangent method is used to determine these values. The R phase start temperature ($R_s$) is the temperature point when the austenite or martensite phase starts to transform into the R phase. The R phase finish temperature ($R_f$) is the temperature point when the transformation of austenite or martensite phase into the R phase is completed. The martensite start temperature ($M_s$) and martensite finish temperature ($M_f$) are the start and finish temperatures of the transformation from R phase to the martensite phase, respectively. Similarly, the austenite start temperature ($A_s$) and austenite finish temperature ($A_f$) are the start and finish temperatures of the transformation from R phase to the austenite phase, respectively. The detailed transformation steps and characteristic temperatures are listed in Table 2. As shown in Fig. 9, the graph presents the typical two transformation steps in the heating and cooling process, which is corresponding to the austenite phase (B2) ↔ R phase ↔ martensite phase (B19'). The occurrence of this phenomenon is related to the $Ni_4Ti_3$ particles appearing in the NiTi matrix, as shown in Fig. 5(a). The fine particles have a strong resistance to the phase transformation with large strain such as B2 ↔ B19', but have much less resistance to the phase transformation with a small strain like B2 ↔ R. The local inhomogeneity in stress and composition promotes the R phase separation from B2 ↔ B19'.

**Table 2**

Characteristic transformation steps and temperatures of raw powders.

| Martensitic transformation (cooling) | | | | Austenitic transformation (heating) | | | |
|---|---|---|---|---|---|---|---|
| Austenite → R phase | | R phase → Martensite | | Martensite → R phase | | R phase → Austenite | |
| $R_s$ (°C) | $R_f$ (°C) | $M_s$ (°C) | $M_f$ (°C) | $R_s$ (°C) | $R_f$ (°C) | $A_s$ (°C) | $A_f$ (°C) |
| 4.5 | -12.4 | -20.6 | -55.3 | -43.5 | -22.4 | -11.5 | 22.9 |

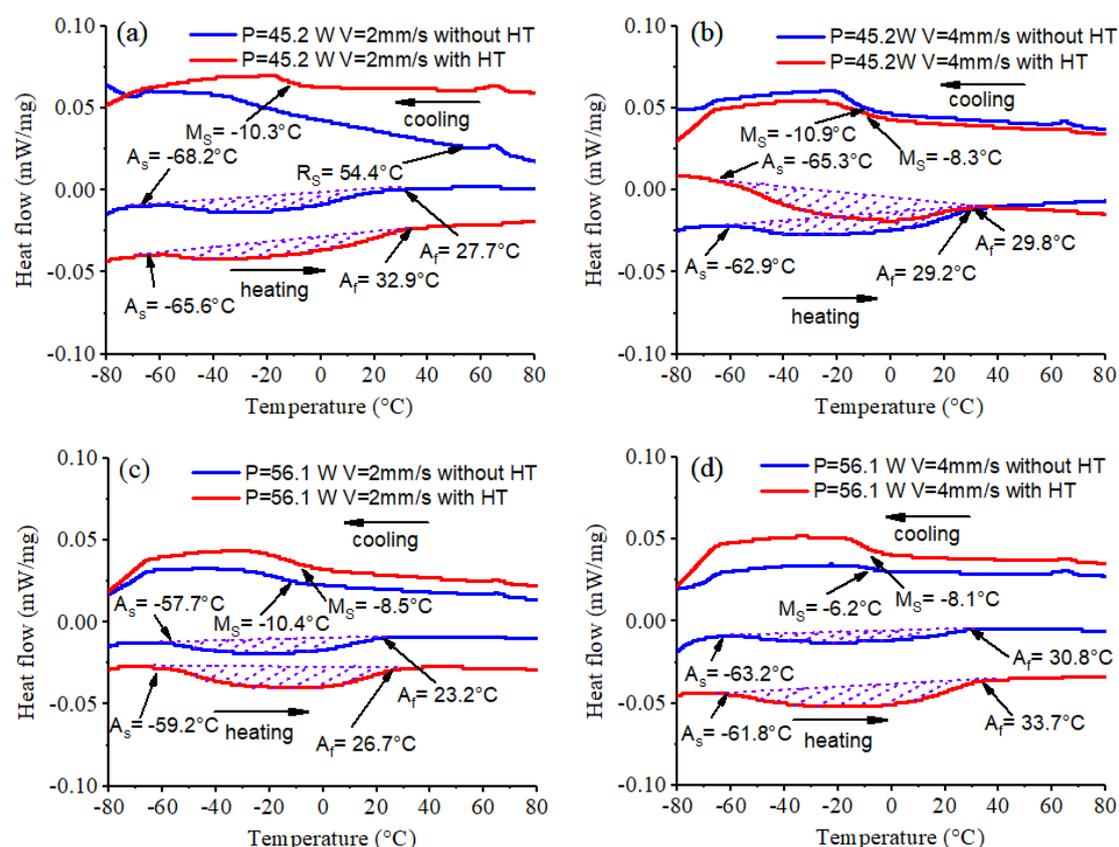

Fig. 10. DSC curves of as-deposited specimens under different process parameters: (a) #1, #5; (b) #3, #7; (c) #2, #6; (d) #4, #8.

The DSC curves for the as-deposited specimens with and without in-situ heat treatment strategy are plotted in Fig. 10. In these figures, the martensite start temperature ($M_s$) is the temperature at which the austenite phase starts to transform into the martensite phase. The R phase start temperature ($R_s$) is the temperature point where the austenite phase starts to transfer into the R phase. The austenite start temperature ($A_s$) and austenite-finish-temperature ($A_f$) are the start and finish temperatures of the transformation from martensite to austenite, respectively. The detailed characteristic transformation temperatures are listed in Table 3. As shown in Fig. 10, a broad phase transformation temperatures (TTs) range is observed in all curves. This relatively wide and flat TTs range is attributed to the formation of microstructure inhomogeneity and chemical segregation in the fabrication process. Due to the fast melting and solidification rate, nonuniform temperature distribution exists between the melting zones and previously deposited layers, inducing residual stress in the fabricated structure. The defects such as lattice dislocation and distortion are then inclined to occur and result in phase transformation resistance. Moreover, chemical segregation is prone to occur in the deposition process. For example, due to the negative thermocapillary coefficient, the Marangoni flow in the molten pool may push Ni solute to the edge and bottom areas of the molten pool. When the temperature drops to the solidification line, the chemical segregation is accompanied by the secondary phase precipitation, leading to gradual and suppressed phase transformation behaviors [56, 57].

**Table 3**

Characteristic transformation steps and temperatures of fabricated samples.

|  | Martensitic transformation (cooling) | | Austenitic transformation (heating) | | |
| --- | --- | --- | --- | --- | --- |
|  | Austenite → Martensite | Austenite → R phase | Martensite → Austenite | | Range |
|  | $M_s$ (°C) | $R_s$ (°C) | $A_s$ (°C) | $A_f$ (°C) | $A_f - A_s$ (°C) |
| #1 | / | 54.4 | -65.6 | 32.9 | 98.5 |
| #5 | -10.3 | / | -68.2 | 27.7 | 95.9 |
| #3 | -8.3 | / | -65.3 | 29.8 | 95.1 |
| #7 | -10.9 | / | -62.9 | 29.2 | 92.1 |
| #2 | -8.5 | / | -57.7 | 23.2 | 80.9 |
| #6 | -10.4 | / | -59.2 | 26.7 | 85.9 |
| #4 | -8.1 | / | -63.2 | 30.8 | 94 |
| #8 | -6.2 | / | -61.8 | 33.7 | 95.5 |

Compared with the DSC curves of the as-deposited sample without in-situ HT, the DSC curves with in-situ HT demonstrate a more obvious phase transformation peak as shown in Fig. 10. Limited by the cooling capacity of the equipment, the cooling curve data for temperatures below -65°C are excluded from our analysis. Hence, the phase transformation temperature ranges for the samples with and without in-situ heat treatment have no obvious difference. The released phase transformation energy contained in the transformation peak presented by the dotted area is highly correlated with the shape memory effect. Li et al. [33] indicated that the smaller energy release is caused by insufficient and inhibited martensite transformation. The improved transformation behaviors of as-deposited samples with in-situ HT are attributed to several factors. The main reason is that the $Ti_2Ni$ phase partially dissolves back to the NiTi (B2) matrix through a reverse peritectic reaction. The local stress caused by the occurrence of precipitation on the grain boundary is thus alleviated. On the other hand, the in-situ HT will result in a slower cooling rate and reduce the thermal gradient throughout the process. The thermodynamic and kinetic factors will become more uniform. The grain inclination towards the scanning direction then will be further limited. The microstructure becomes more uniform, weakening lattice dislocation. The reduction of $Ti_2Ni$ precipitation and the improvement in microstructure homogeneity all contribute to a more obvious phase transformation peak in the DSC curves for samples fabricated with in-situ HT.

## 3.4 Microhardness

The microhardness graphs of fabricated NiTi alloys with and without in-situ HT are presented in Fig. 11. These curves do not show a specific trend or variation for different process parameters. This phenomenon is highly related to the overall performance of the microstructure of as-deposited specimens. Firstly, the phase constituents retained in the specimens have a huge influence on the microhardness since the austenite phase (B2) and secondary phase ($Ti_2Ni$) are harder than the martensite phase (B19'). Secondly, the larger grain size and fabrication defects will decrease the microhardness and vice versa. The effect of grain size on the mechanical properties of fabricated samples can be explained by the Hall-Petch relationship [58]. The final value of measured microhardness is determined by the combined effect of grain size, defects, and remaining phases.

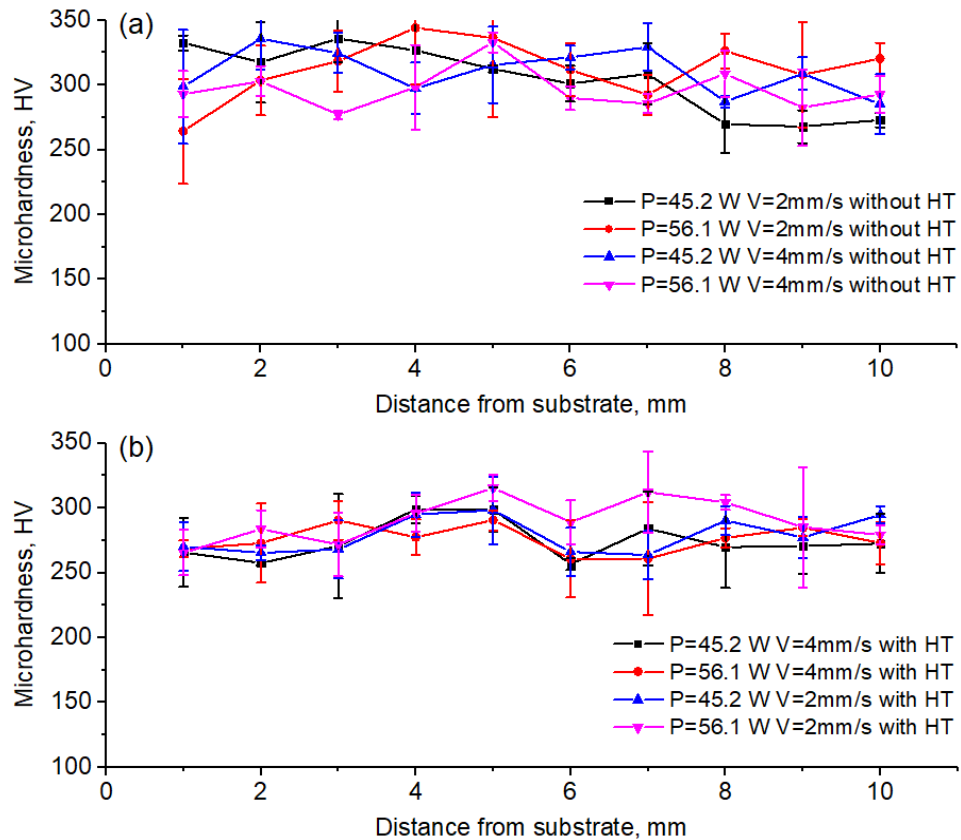

Fig. 11. Microhardness of as-deposited specimens at the longitudinal cross-section with and without HT (load: 100 g, hold:15 s).

By comparing Fig. 11(a) and Fig. 11(b), it can be seen that the microhardness of as-deposited samples with HT strategy is lower than the microhardness of fabricated specimens without HT strategy. An approximate 35 HV reduction in microhardness is detected after in-situ HT. Since the grain sizes of as-deposited samples with and without HT have no obvious difference, as shown in Fig. 12. At the same time, the manufacturing defects are also similar for the fabricated samples with and without HT, as shown in Fig. 6 and Fig. 7. The microhardness reduction can mainly be attributed to the reduction of the $Ti_2Ni$ phase, which has a higher hardness in comparison with the NiTi matrix [59].

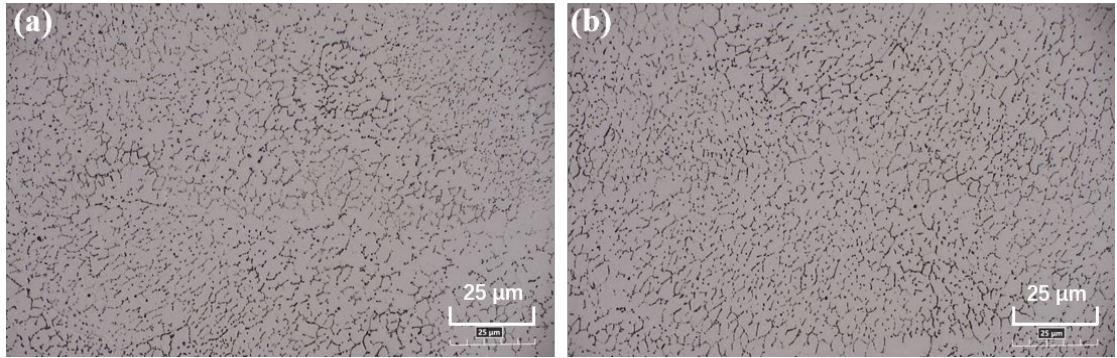

Fig. 12. Microstructure of as-deposited samples fabricated with and without in-situ HT: (a) P = 45.2 W, V = 2 mm/s without HT, (b) P = 45.2 W, V = 2 mm/s with HT.

## 4. Conclusion

In this study, an in-situ HT strategy is adopted to assist the fabrication of NiTi shape memory alloys. A low-power laser beam scans the same path after each layer of energy deposition to reheat the last deposited layer to achieve a reverse peritectic reaction and a transient high solution treatment. The in-situ HT helps to dissolve the $Ti_2Ni$ phase back into the NiTi matrix as well as homogenize the microstructures, thereby potentially achieving the spatial control of thermal and mechanical properties of fabricated NiTi parts. The effectiveness of in-situ HT is verified by investigating the temperature histories using calibrated simulations and comparing experimental characterizations in terms of XRD patterns, phase compositions, DSC curves, and microhardness.

The 3D FE simulation results indicate that the temperature evolution of in-situ HT reaches around 1200°C, above the peritectic reaction temperature of the $Ti_2Ni$ phase. The top layer during each reheating process will then experience a reverse peritectic reaction and a transient high solution treatment successively. The microstructure analysis shows that the in-situ HT method helps to reduce the $Ti_2Ni$ phase in the NiTi matrix by 50%–70% compared with the samples fabricated without in-situ HT. The decrease of the $Ti_2Ni$ phase is ascribed to the partial dissolution of $Ti_2Ni$ back into the NiTi matrix. The DSC measurement results show that the in-situ HT method improves the shape memory effect by reducing the $Ti_2Ni$ precipitation and improving microstructure homogeneity during the deposition. The microhardness investigation shows a reduction of 35 HV by in-situ HT due to the reduction of the $Ti_2Ni$ phase.

It is demonstrated that the proposed in-situ HT strategy can help to improve the transformation temperature and microstructure of deposited NiTi alloys. By adopting a dynamically adjusted laser power and adaptive tool path trajectories, it is possible to extend this method to achieve localized heat treatment for the spatial control of thermal and/or mechanical properties.

## Acknowledgments

This research was supported by the start-up fund from McCormick School of Engineering, Northwestern University, Evanston, IL, USA; the Research Grants Council of the Hong Kong Special Administrative Region, China (Project No. CUHK 14202219) and The Chinese University of Hong Kong (Project ID: 4055117).

**Appendix. Simulation and experiment details**

Pre-alloyed $Ni_{50.93}Ti_{49.07}$ (at.%) powders, supplied by AMC Powders (China), were used as the precursor material. The powder had spherical or near-spherical morphology with a particle size range from 20 to 50 μm. A NiTi plate with similar chemical compositions was employed as the substrate to improve binding and minimize warpage. A customized DED setup equipped with a 500W multi-mode laser source (YLR-500-MM-AC-Y14, *IPG Photonics*), a 6-axis parallel platform (H840.D11, *PI*), a co-axial ring nozzle (COAX 40-F, *Fraunhofer ILT*), and a powder feeder (GPV PF2/2, *GTV Thermal Spray*) was employed to fabricate the NiTi parts as shown in Fig. A1. The whole deposition process was conducted in an inert gas atmosphere, where the oxygen and water content was controlled to be less than 100 ppm.

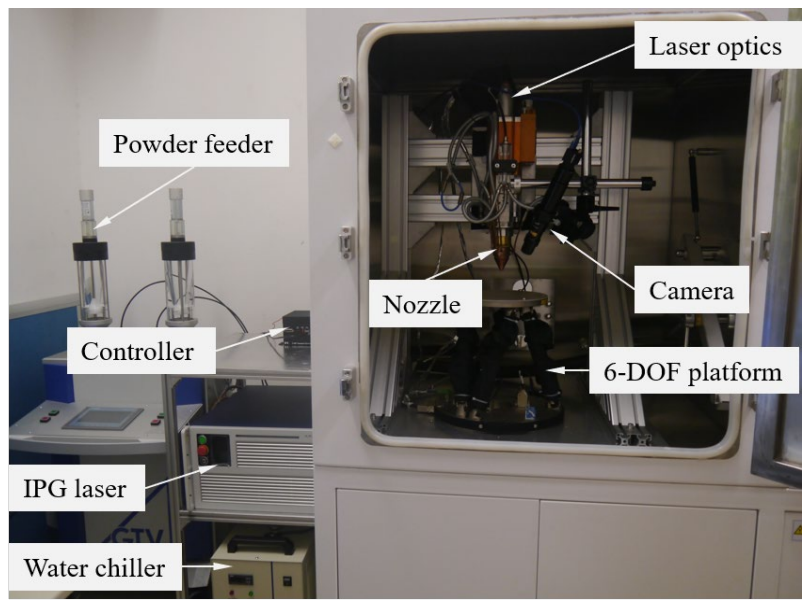

Fig. A1. Photo of directed energy deposition setup.

The simulation model geometry was set according to the experimental condition. For model accuracy and efficiency, the four-node tetrahedron element was assigned to the substrate domain while the eight-node hexahedron element was assigned to the deposited structure domain. For each single-wall layer, there were four elements set in the depth direction and six elements set in the width direction. The element size ranged from 7.5 μm–12.5 μm in the depth direction and 50 μm in the width and length directions. A similar mesh density was adopted in previous research [60-62]. The PARDISO solver with a fully coupled solution was utilized to perform the calculation. The movement of the laser trajectory was controlled by a user-defined subroutine. The substrate temperature was measured by a thermocouple, which was employed to calibrate the model parameters as described in Section 2.3. The temperature-dependent material properties of pre-alloyed NiTi ($Ni_{50.93}Ti_{49.07}$ at.%) were adopted for the simulation model, which was calculated using JMatPro software based on the phase diagrams. The power effective absorption coefficients were determined by the calorimetric method using the data in our previous experiment [63]. The detailed model parameters are given in Fig. A2 and Table A1-A2.

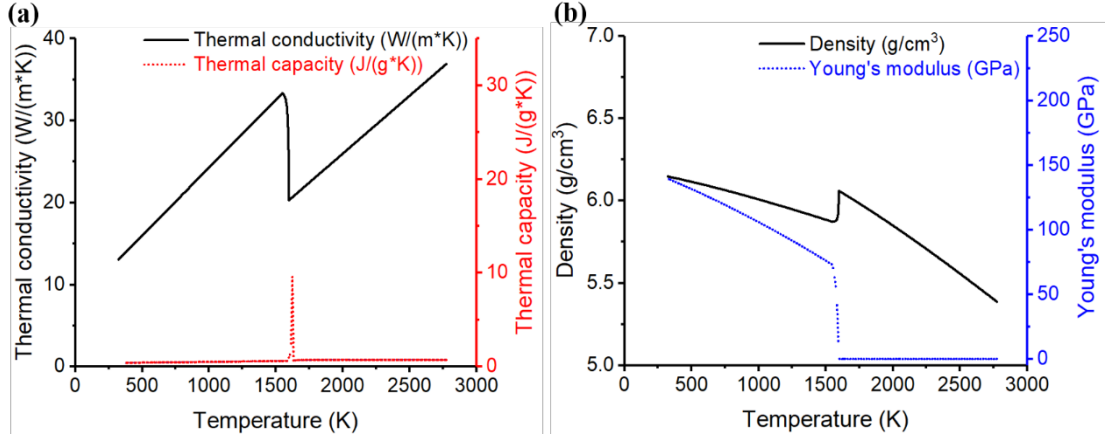

Fig. A2. Thermal-dependent material properties of NiTi alloy: (a) conductivity and capacity; and (b) density and Young's modulus.

**Table A1**

Material properties used in the simulation.

| Parameters | Values |
| --- | --- |
| Density, $\rho$ | T-dep, 5.4-6.45 g/cm$^3$ |
| Thermal conductivity, $k$ | T-dep, 13-36 W/(m*K) |
| Specific heat capacity, $c_p$ | T-dep, 470-840 J/(kg*K) |
| Thermal expansion coefficient, $\alpha$ | 11e-6, 1/K |
| Solidus temperature, $T_s$ | 1553 K |
| Liquidus temperature, $T_l$ | 1583 K |
| Latent heat, $L_m$ | 24.2 kJ/kg |

**Table A2**

Deposition parameters used in the simulation.

| Parameters | #1, #5 | #2, #6 | #3, #7 | #4, #8 |
| --- | --- | --- | --- | --- |
| Power effective absorption coefficient, $\eta$ | 0.293 | 0.311 | 0.287 | 0.278 |
| Laser power, $P$ (W) | 45.2 | 56.1 | 45.2 | 56.1 |
| Scan speed, $v_p$ (mm/s) | 2 | 2 | 4 | 4 |
| Width of the molten pool, $a$ (μm) | 185 | 228 | 157.5 | 211.5 |
| Depth of the molten pool, $b$ (μm) | 46 | 63 | 40 | 50 |
| Front length of the molten pool, $c_f$ (μm) | 195 | 234 | 167.5 | 221.5 |
| Rear length of the molten pool, $c_r$ (μm) | 254.7 | 272 | 173 | 265 |